\documentclass[useAMS,usenatbib,epsfig, a4paper,amsmath,aps]{mnras}
 
\usepackage{float}
\usepackage{epsfig}
\usepackage{subfig}
\usepackage{verbatim}
\usepackage{nicefrac}
\usepackage{url}
\usepackage{mathrsfs}
\usepackage{amssymb}
\usepackage{amsmath}
\usepackage{color}

\newcommand{\deltamu}{\Delta \mu}

\title[Discerning Dark Energy Models with High-Redshift Standard Candles]
   {Discerning Dark Energy Models with High-Redshift Standard Candles}
\author[P.~Andersen et al.]
{P.~Andersen$^{1}$\thanks{email: \href{mailto:perandersen@dark-cosmology.dk}{\nolinkurl{perandersen@dark-cosmology.dk}}} and J. Hjorth$^{1}$\\ 
$^{1}$Dark Cosmology Centre, University of Copenhagen, Juliane Maries Vej 30, 2100 Copenhagen O, Denmark.
}

\date{\today}
\pubyear{2016}

\begin{document}
\maketitle

\begin{abstract}
Following the success of type Ia supernovae in constraining cosmologies at lower redshift $(z\lesssim2)$, effort has been spent determining if a similarly useful standardisable candle can be found at higher redshift. {In this work we determine the largest possible magnitude discrepancy between a constant dark energy $\Lambda$CDM cosmology and a cosmology in which the equation of state $w(z)$ of dark energy is a function of redshift for high redshift standard candles $(z\gtrsim2)$}. We discuss a number of popular parametrisations of $w(z)$ with two free parameters, $w_z$CDM cosmologies, including the Chevallier-Polarski-Linder and generalisation thereof, $n$CPL, as well as the Jassal-Bagla-Padmanabhan parametrisation. For each of these parametrisations we calculate and find extrema of $\deltamu$, the difference between the distance modulus of a $w_z$CDM cosmology and a fiducial $\Lambda$CDM cosmology as a function of redshift, given 68\% likelihood constraints on the parameters $P=(\Omega_{m,0}, w_0, w_a)$. The parameters are constrained using cosmic microwave background, baryon acoustic oscillations, and type Ia supernovae data using CosmoMC. We find that none of the tested cosmologies can deviate more than 0.05 mag from the fiducial $\Lambda$CDM cosmology at high redshift, implying that high redshift standard candles will not aid in discerning between a $w_z$CDM cosmology and the fiducial $\Lambda$CDM cosmology. Conversely, this implies that if high redshift standard candles are found to be in disagreement with $\Lambda$CDM at high redshift, then this is a problem not only for $\Lambda$CDM but for the entire family of $w_z$CDM cosmologies.
\end{abstract}


\section{Introduction}
\label{sec:intro}
The concordance $\Lambda$CDM model, containing a dark energy ($\Lambda$) component with constant equation of state and a cold dark matter (CDM) component, has been successful in explaining observations of a large number of cosmological probes, including supernovae (SNe), baryon acoustic oscillations (BAO) and the power spectrum of the cosmic microwave background (CMB). Anomalies do however exist \citep{PlanckCollaboration2015}, and a number of alternatives to $\Lambda$CDM have been proposed.  One proposed modification is allowing the equation of state to vary with redshift.  We call this family of models $w_z$CDM models. A number of other alternative modifications include modified gravity, such as $f(R)$ models \citep{Buchdahl1970, Sotiriou2008} where the Ricci scalar $R$ is replaced with a function of $R$, or redshift remapping \citep{Wojtak2016} where the assumption of a $a = (1+z)^{-1}$ relation between the scale factor $a$ and redshift $z$ is broken in favour for a relation that is a function of redshift. In this work we focus exclusively on the first kind of models, the $w_z$CDM cosmologies.\\

Type Ia supernovae (SNe) are used as standard candles at lower redshifts. At redshifts of $z \gtrsim 2$ they are less useful, in part due to the decreasing Ia SNe rate \citep{Rodney2014}. Recently a range of high redshift standard candles have been proposed, including active galactic nuclei (AGN) \citep{king2013,Honig2016}, gamma ray bursts (GRB) \citep{Amati2016}, gamma ray burst supernovae (GRB-SNe) \citep{li2014, Cano2014}, superluminous supernovae (SLSNe) \citep{Inserra2014, scovacricchi2016}, quasars \citep{Risaliti2015, Lopez-Corredoira2016} and high redshift \ion{H}{II} galaxies \citep{Terlevich2015}. In this work we investigate the usefulness of such high redshift standard candles for constraining dark energy models. We choose to neither use a Fisher matrix approach, where $\Lambda$CDM parameters are assumed, nor to assume a $\Lambda$CDM cosmology to generate mock datasets. Rather, we introduce the quantity $\deltamu(z)$ which is defined as the difference between the distance modulus of a $w_z$CDM cosmology and fiducial $\Lambda$CDM cosmology as a function of redshift. Fiducial $\Lambda$CDM cosmology is in this work defined as the best fitting $\Lambda$CDM cosmology. Previous work has applied a similar approach \citep{perlmutter2003, Aldering2007} to argue that discerning a non-constant dark energy component from a constant dark energy will require high precision measurements. Since then the amount of data to constrain proposed cosmologies has grown, not only in quantity but also extended to increasingly higher redshifts. This allows us to revisit this approach, specifically asking whether any $w_z$CDM cosmology can deviate significantly in predicted distance modulus from fiducial $\Lambda$CDM at high redshift, given current cosmological constraints. If $w_z$CDM cosmologies are indistinguishable from the fiducial $\Lambda$CDM cosmology at high redshifts, this would imply that high redshift $(z\gtrsim2)$ data points will have limited usefulness over low redshift $(z\lesssim2)$ equivalents when discerning between the two types of cosmologies. Additionally it would imply that if precise measurements of standard candles at high redshift are shown to be in disagreement with $\Lambda$CDM cosmology, this would not only challenge $\Lambda$CDM cosmology but also the entire family of $w_z$CDM cosmologies.\\

Quintessence is a proposed form of dark energy \citep{Ratra1988, Caldwell1998, Tsujikawa2013a}, which introduces a scalar field that is minimally coupled to gravity to explain the apparent accelerated expansion observed at low redshifts. In this work we use theoretical concepts from quintessence to guide us in determining which $w_z$CDM models to test. In quintessence exists two subgroups of dark energy models, the thawing \citep{Scherrer2008, Chiba2009, Gupta2015} and the freezing \citep{Scherrer2006, Sahlen2007, Schimd2007} models. In thawing dark energy models the scalar field is nearly frozen due to the potential being dampened by Hubble friction in the early matter dominated universe, with the scalar field then starting to evolve once the field mass is below the Hubble expansion rate. In freezing dark energy models the potential is steep enough in the early universe that a kinetic term develops. At later stages the evolution of the field, and therefore also the evolution of the equation of state, steadily slows down as the potential is tending towards being shallow. These two families of models produce distinct behaviours for the evolutions of the equation of state. The equation of state function $w(z)$ of thawing dark energy models are generically convex decreasing functions of redshift, while freezing dark energy models produce $w(z)$ functions that are convex increasing functions of redshift. Phenomenological parametrisations of $w(z)$ with two free parameters can produce either convex increasing or decreasing behaviour, but not both, making them more suited to fit either an underlying freezing or thawing dark energy model \citep{pantazis2016}.\\

The $w(z)$ parametrisations investigated in this work will include some generalisations of $\Lambda$CDM, limited to two parameters $w_0$ and $w_a$ where $w_0$ is a constant term and the magnitude of $w_a$ determines the strength of the evolution with redshift. Specifically we investigate the Chevallier-Polarski-Linder (CPL) \citep{chevallier2000, linder2002} and Jassal-Bagla-Padmanabhan (JBP) \citep{jassal2004} parametrisations as well as the $n$CPL generalisation of \cite{pantazis2016}. \\

In section \ref{sec:deltaparameter} we introduce and motivate the parameter $\deltamu$, and discuss previous work that has used a similar approach. Then in section \ref{sec:parametrisations} we discuss parametrisations of $w(z)$ and the reasoning behind choosing the subset of parametrisations adopted in this paper. In section \ref{sec:method}, we describe the method used to derive extrema of $\deltamu$ for the chosen parametrisations, and in section \ref{sec:results} we present the results. Finally in section \ref{sec:discussion} we discuss the implications of our results for using high redshift standard candles to constrain dark energy models.\\

\section{The $\deltamu$ parameter}
\label{sec:deltaparameter}
Previous works have discussed the utility of high redshift standard candles in constraining cosmological parameters \citep{king2013, scovacricchi2016} by using either Fisher matrix formalism or simulating data from a high redshift standard candle. While these approaches are appropriate for forecasting the constraining power of a survey, they are less suited for our purpose. Therefore we introduce the $\deltamu$ parameter, given as $\deltamu(z,P) = \mu_{w_{z}\mathrm{CDM}}(z,P) - \mu_{\Lambda \mathrm{CDM}}(z)$. Here $\mu_{w_{z}\mathrm{CDM}}(z,P)$ is the distance modulus of a $w_z$CDM cosmology given a set of parameters $P=(\Omega_{m,0}, w_0, w_a)$, and $\mu_{\Lambda \mathrm{CDM}}(z)$ the distance modulus of the best fitting $\Lambda$CDM cosmology, for a given redshift $z$.\\

Looking at the differences in magnitude as a function of redshift predicted by different cosmologies has been done before in the literature. \cite{perlmutter2003} plot the magnitude difference between a ($\Omega_{m,0}$, $\Omega_{\Lambda,0}$)=(0.3, 0.7) $\Lambda$CDM cosmology and a $w$CDM cosmology where the equation of state of dark energy is allowed to vary between $w=-1.5$ and $w=-0.5$, out to a redshift of $z=1.7$. With the parameter range they find a maximum magnitude difference of approximately $\sim$0.3 mag. \cite{Aldering2007} determine the magnitude difference between a $\Lambda$CDM cosmology with $w=-1$ and $w$CDM cosmologies with $w=+1$, $w=0$, and $w=-\infty$, respectively, in the redshift range [1.7,3]. The maximum magnitude difference they find is less than 0.04 mag. In this work we define a method that for a chosen $w_z$CDM cosmology selects a range of parameter values $P=(\Omega_{m,0}, w_0, w_a)$ which reflect how strongly constrained the parameters are by the data. This is done by selecting all parameter values that lie within the 68\% likelihood contour spanning the $\Omega_{m,0}$, $w_0$, and $w_a$ parameter space, given observational constraints. The method is described in detail in section \ref{sec:method} and Appendix \ref{ap:extremevaluetheorem}.

\section{Dark Energy Equation of State Parametrisations}
\label{sec:parametrisations}
\begin{figure}
  \centering\includegraphics[width=\linewidth]{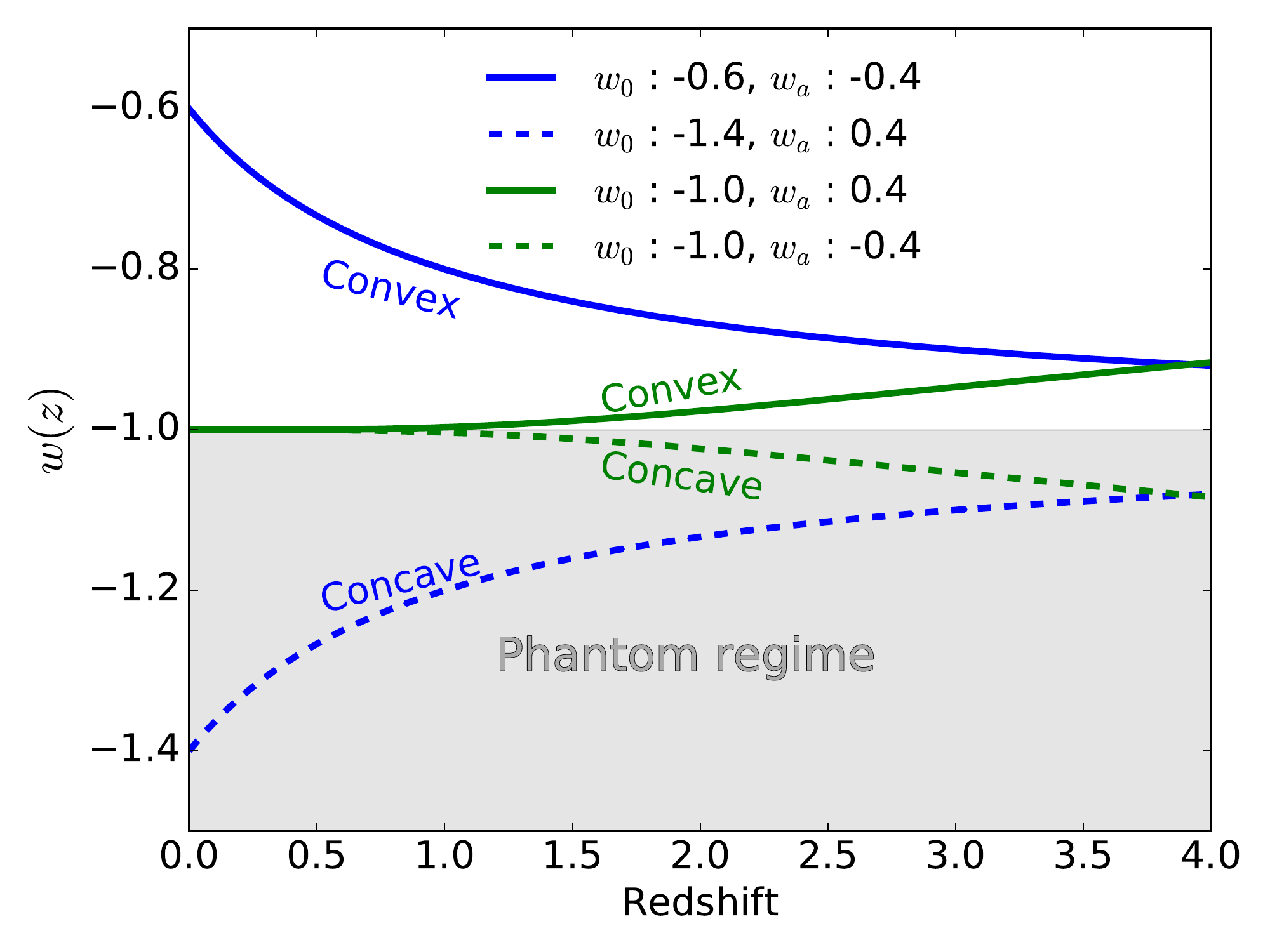}
  \caption{A plot of the equation of state parameter as a function of redshift for the CPL (blue) model and $n7$CPL (green) model. Looking at the full lines this figure illustrates the convex decreasing and increasing behaviour of thawing and freezing dark energy models, respectively. If the $w(z)$ are allowed to enter the phantom regime where $w(z)<-1$ then the dashed concave behaviour becomes possible. In the phantom regime we see the CPL function being concave instead of convex, and increasing instead of decreasing. Likewise the $n7$CPL model in the phantom regime changes behaviour to become concave instead of convex, and decreasing instead of increasing. }
  \label{fig:equationofstate}
\end{figure}
Testing all proposed models for $w(z)$ is not possible so a sample thereof must be chosen. We use concepts from quintessence dark energy models as a guide. Specifically we apply the predicted evolution of $w(z)$ from the two subclasses of quintessence, the thawing and freezing models. In quintessence, thawing models produce convex decreasing $w(z)$ functions of redshift whereas freezing models produce convex increasing $w(z)$ functions of redshift. When going beyond thawing and freezing quintessence, phenomenological models can produce concave behaviour of $w(z)$ and also enter the phantom regime where $w(z)<-1$. This is illustrated in Fig.~\ref{fig:equationofstate}. In this work we do not exclude solutions where the $w(z)$ function enters the phantom regime. However, we clearly indicate any results in both figures and discussion that arise from such phantom models.\\

Regardless of whether the phantom regime is excluded, models better suited for fitting an underlying thawing cosmology are unable to deviate significantly from the fiducial $w=-1$ at high redshifts unless they also do so at low redshifts. The bulk of the current cosmological constraints exists at low redshifts. We therefore expect that models better suited for fitting an underlying freezing dark energy are able to deviate from fiducial $\Lambda$CDM cosmology at higher redshifts more than models better suited for fitting an underlying thawing dark energy model. This motivates dividing phenomenological models into two categories, those better suited for fitting an underlying thawing dark energy and those better suited for fitting and underlying freezing dark energy. \cite{pantazis2016} show that CPL, JBP, and $n$=3 $n$CPL ($n3$CPL) are better suited for fitting an underlying thawing cosmology and a $n$=7 $n$CPL ($n7$CPL) model is better suited for fitting an underlying freezing cosmology.\\

A phenomenological model better suited for fitting an underlying thawing dark energy can reproduce observables for a cosmology with a freezing dark energy. \cite{deputter2008} show that the CPL model reproduces distances of freezing dark energy models, such as the supergravity inspired SUGRA model, to within 0.1\%. However, as shown by \cite{pantazis2016}, fitting a freezing dark energy with a model better suited for fitting an underlying thawing dark energy, and vice versa, can lead to incorrect values of $w(z=0)$ or phantom behaviour which are not present in the underlying cosmology.\\

We therefore include the CPL model \citep{chevallier2000, linder2002}
\begin{equation}
w_{\mathrm{CPL}} = w_0 + w_a(1-a) = w_0 + w_a \frac{z}{1+z},
\end{equation}
the JBP model \citep{jassal2004}
\begin{equation}
w_{\mathrm{JBP}} = w_0 + w_a(1-a)a = w_0 + w_a \frac{z}{(1+z)^2},
\end{equation}
and the $n$CPL model \citep{pantazis2016}
\begin{equation}
w_{n\mathrm{CPL}} = w_0 + w_a(1-a)^n = w_0 + w_a \left(\frac{z}{1+z}\right)^n,
\end{equation}
where we, guided by Fig.~16 of \cite{pantazis2016}, choose a $n3$CPL and a $n7$CPL cosmology. Since models appropriate for fitting an underlying freezing dark energy are much less constrained in especially the $w_a$ parameter, running CosmoMC until convergence takes much longer time for this type of model than for the models better suited for fitting an underlying thawing dark energy. Therefore only one model of the first type is included, namely the $n7$CPL model.\\

\section{Method}
\label{sec:method}
To derive 68\% likelihood constraints on the models described in the previous section, the CosmoMC \citep{Lewis2002, Lewis2013} MCMC tool was utilised. All cosmologies were fit with the same data, including observations of type Ia supernovae from the Joint Light-Curve Analysis, baryon acoustic oscillations from SDSS-III and 6dF and the cosmic microwave background from the Planck Collaboration. For more details and references to the data sets used see Appendix \ref{ap:datanalysis}. To allow maximum flexibility in the fitting routines all parameters for the datasets used were kept free for CosmoMC to fit, e.g. the stretch and colour parameters $\alpha$ and $\beta$ for the JLA sample. Since our focus is to investigate the effects of different dark energy models, and to limit computation time, we assume flatness, and neglect radiation. For each of the models described above (CPL, $n$CPL, and JBP) we derive the 3-dimensional likelihood contours of $\Omega_{m,0}$, $w_0$, and $w_a$. When searching for extrema of $\deltamu(P,z)$ as a function of redshift, only the parameter values $P=(\Omega_{m,0}, w_0, w_a)$ that lie on the surface of the 68\% likelihood volume are used. Appendix \ref{ap:extremevaluetheorem} discusses how this is possible by applying the extreme value theorem.\\

For a Hubble constant in units of km s$^{-1}$Mpc$^{-1}$, the distance modulus is given by
\begin{align}
\label{eq:deltamu}
\begin{split}
\mu(P,z) &= 5\log_{10}{[D_L'(P,z)]}+5\log_{10}{[c/H_0]}+25+\sigma_m,
\end{split}
\end{align}
where $D_L'(P,z)$ is the unitless luminosity distance ($D_L'$=$D_L \,H_0 \, c^{-1}$), $D_L$ is the luminosity distance, {and $\sigma_m$ is a constant representing how far this magnitude is from the correct intrinsic absolute magnitude of the observed supernovae}. $\sigma_m$ and $5\log_{10}{[c/H_0]}$ are both additive terms, and combined in a parameter $\mathit{K}$ to be marginalised over, 
\begin{equation}
\label{eq:constantk}
K=5\log_{10}{[c/H_0]}+25+\sigma_m.
\end{equation}
\\

The marginalisation process is performed by minimising the sum
{
\begin{align}
\label{eq:minimisation}
\chi^2(\mathit{K})&=\sum_i \left( \frac{\mu_{w_z\mathrm{CDM},i}(P,z) - \mu_{\mathrm{JLA},i}}{\sigma_{i,\mathrm{JLA}}} \right)^2\\
			&=\sum_i \left( \frac{5\log_{10}{[D_L'(P,z)_{w_z\mathrm{CDM},i}]} + \mathit{K} - \mu_{\mathrm{JLA},i}}{\sigma_{i,\mathrm{JLA}}} \right)^2
\end{align}
}
by varying $\mathit{K}$. {The sum goes over the JLA type Ia SNe \citep{Betoule2014} with distance modulus $\mu_{\mathrm{JLA},i}$ and associated uncertainty on the distance modulus $\sigma_{i,\mathrm{JLA}}$}. The scheme explained above to recover $\mathit{K}$ is similar in purpose to the process described in Appendix A.1 of \cite{goliath2001}. The process of finding the value of $\mathit{K}$ that minimises Eq.~\ref{eq:minimisation} is done for all parameter values $P=(\Omega_{m,0}, w_0, w_a)$ that lie on the surface of the 68\% likelihood contour. Having determined the value $\mathit{K}$ that minimises Eq.~\ref{eq:minimisation}, we can calculate $\deltamu$ as a function of redshift for parameters $P$ of, e.g., the CPL model
\begin{align}
\begin{split}
&\deltamu_{CPL}(P,z) = \\
&\mu_{\mathrm{CPL}}(P,z) + K_{\mathrm{CPL}}(P) -( \mu_{\Lambda \mathrm{CDM,bf}}(z) + K_{\Lambda \mathrm{CDM},bf}).
\end{split}
\end{align}
In summary the process to derive $\deltamu$ values for a given dark energy model is
\begin{enumerate}
\item Run CosmoMC for given dark energy model.
\item Derive from CosmoMC output 68\% likelihood contours for parameters $P=(\Omega_{m,0}, w_0, w_a)$.
\item Calculate $\mathit{K}$ for all parameter values $P=(\Omega_{m,0}, w_0, w_a)$ on 68\% likelihood contour.
\item Calculate $\deltamu(P,z)$ for all parameter values $P=(\Omega_{m,0}, w_0, w_a)$ on 68\% likelihood contour. 
\end{enumerate}
The above process would not be necessary if $H_0$ and the absolute magnitude of the standard candle were known exactly. If they were known, their values could be substituted directly into Eq.~\ref{eq:deltamu}.
\begin{figure*}
  \centering\includegraphics[width=\linewidth, height=0.78\textheight]{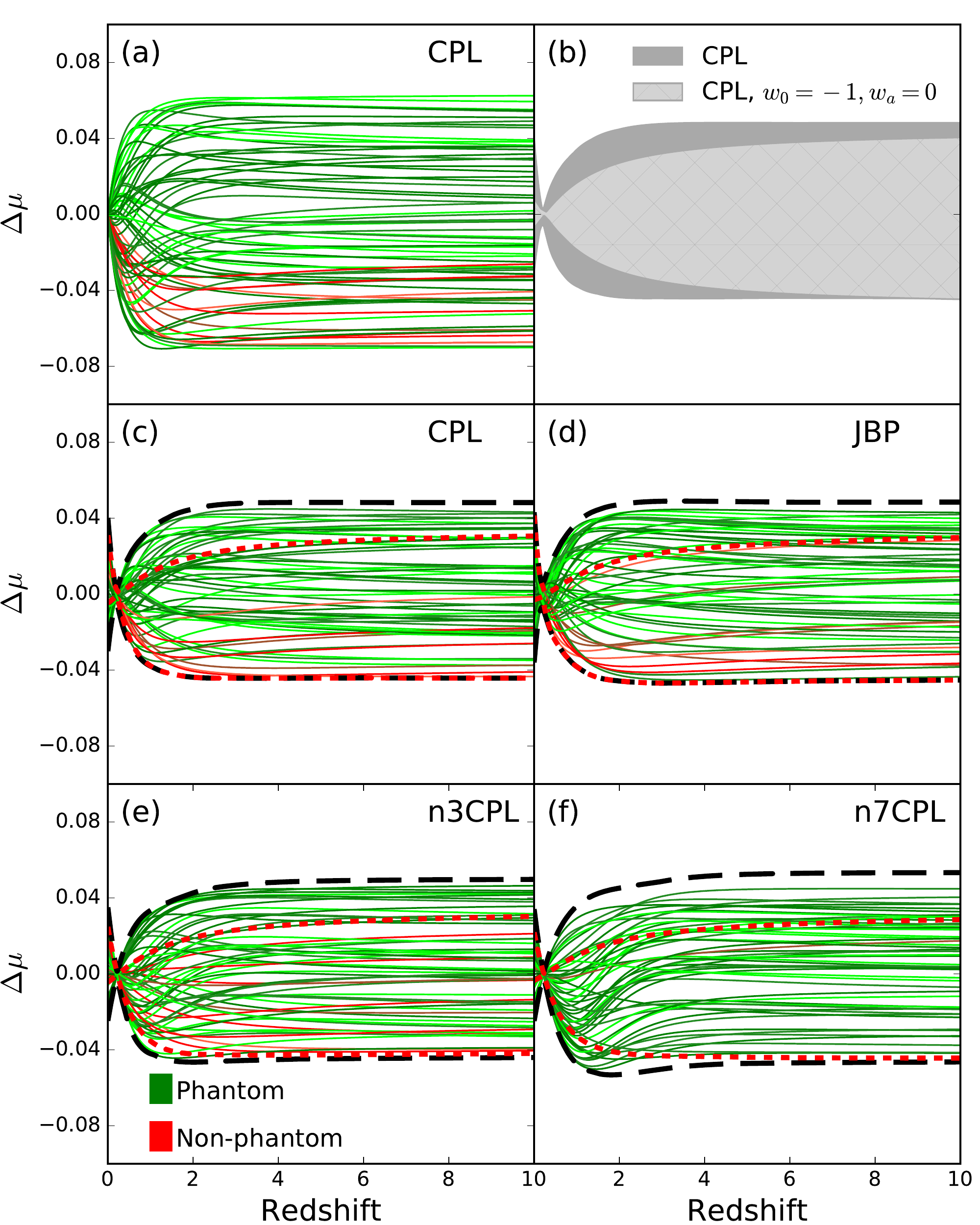}
  \caption{ { $\deltamu(z,P)$ as a function of redshift for the tested cosmologies. The green and red lines show $\deltamu(z,P)$ curves for a representative subsample of all tested parameters $P=(\Omega_{m,0}, w_0, w_a)$. Green lines correspond to $\deltamu$ values from a dark energy model that crosses the phantom divide, $w<-1$, where red lines are models where for all redshifts $w \geq -1$. Likewise the black dashed lines show extrema when placing no restrictions on evolution of equation of state and red dashed lines showing extrema when excluding models where the equation of state enters the phantom region. {\bf (a)} $\deltamu$ plot for the CPL $w_z$CDM cosmology showing the extrema of $\deltamu(z,P)$ as a function of redshift given the 68\% likelihood constraints derived using CosmoMC, ignoring the marginalisation constant $K$ from Eq.~\ref{eq:constantk}. {\bf (b)} $\deltamu$ plot for the CPL $w_z$CDM cosmology, showing the extrema of $\deltamu(z,P)$ as a function of redshift, given the 68\% likelihood constraints derived using CosmoMC. The dark grey shaded area is where all parameters $\Omega_{m,0}$, $w_0$, $w_a$ have been allowed to vary, and the light grey hatched where only $\Omega_{m,0}$ varies and $w_0$ and $w_a$ are fixed at the $\Lambda$CDM values of $w_0=-1$ and $w_a=0$ respectively. The plot illustrates that most of the magnitude discrepancy with the best fitting $\Lambda$CDM comes from the uncertainty in $\Omega_{m,0}$, rather than the choice of dark energy model and uncertainty in $w_0$ and $w_a$. {\bf (c)} $\deltamu$ plot for the CPL $w_z$CDM cosmology, showing the extrema of $\deltamu(z,P)$ as a function of redshift, given the 68\% likelihood constraints derived using CosmoMC. Black dashed lines show extrema for $\deltamu$. Red dashed lines show extrema for $\deltamu$ excluding dark energy models that at any redshift has a phantom value for the quation of state of dark energy, $w<-1$. {\bf (d)} Like panel (c) but for the JBP $w_z$CDM cosmology. {\bf (e)} Like panel (c) but for the n3CPL $w_z$CDM cosmology. {\bf (f)} Like panel (c) but for the n7CPL $w_z$CDM cosmology.} }
  \label{fig:combinedplot}
\end{figure*}

\section{Results}
\label{sec:results}
The results are shown in Fig.~\ref{fig:combinedplot}. Before discussing them we address two important topics. Firstly, the effects of the constant $K$ (Eq.~\ref{eq:constantk}). Secondly, how large a part of $\deltamu$ that originates from uncertainty in $\Omega_{m,0}$ and how large a part that comes from the choice of dark energy model and uncertainty in $w_0$ and $w_a$. In discussing these topics we only investigate the CPL model in detail and summarise results for the JBP, $n3$CPL, and $n7$CPL models; an in-depth discussion for the latter three models can be found in Appendix \ref{ap:additionalplots}.

\subsection{Effects of $K$}
\label{sec:effectsofk}
$\deltamu$ values for a representative sample of the parameters $P=(\Omega_{m,0}, w_0, w_a)$ lying on the 68\% likelihood surface of $\Omega_{m,0}$, $w_0$, $w_a$ parameter space for the the CPL cosmology are shown in panel (a) of Fig.~\ref{fig:combinedplot}. Importantly in this figure the marginalisation constant $K$ from Eq.~\ref{eq:constantk} has been ignored. As one would then expect, $\deltamu \rightarrow 0$ for $z \rightarrow 0$. However, ignoring the marginalisation constant $K$ gives an incomplete picture. Neither the Hubble constant nor the intrinsic absolute magnitude of the type Ia SNe are known precisely. If they were then panel (a) would be appropriate, but since they are not we must include the marginalisation parameter $K$. The $\deltamu$ values for the CPL cosmology, including the marginalisation constant $K$, are shown in panel (c) of Fig.~\ref{fig:combinedplot}, with the extrema of $\deltamu$ shown as dashed lines. Including the marginalisation constant introduces scatter around $z \approx 0$, which is due to the fact that the values of $K$ differ for different parameter values. Furthermore, the extrema of the $\deltamu$ values become smaller. This is to be expected, as the marginalisation process finds the value of $K$ that minimises the difference between the distance moduli of the cosmological models and the observed SNe Ia magnitudes, for both the $w_z$CDM and the $\Lambda$CDM cosmology.  This decreases any disagreement that might exist between the predicted distance moduli of $w_z$CDM and $\Lambda$CDM cosmology. This result also holds true for the JBP, $n3$CPL, and $n7$CPL cosmologies, i.e. including $K$ causes scatter around $z=0$ and a narrower distribution of $\deltamu$ values.\\

\subsection{$\deltamu$ contribution from $\Omega_{m,0}$ versus $w_0$ and $w_a$}
\label{sec:contributions}
When looking at the extrema of $\deltamu$ as a function of redshift it is not straightforward to disentangle how large a part of the magnitude difference is caused by uncertainty in $\Omega_{m,0}$ and how much stems from the choice of CPL $w_z$CDM cosmology and associated uncertainty in $w_0$ and $w_a$. Therefore we produce panel (b) of Fig.~\ref{fig:combinedplot}. The dark grey shaded area in panel (b) of Fig.~\ref{fig:combinedplot} is the range of possible $\deltamu$ values when all parameters, $\Omega_{m,0}$, $w_0$, and $w_a$ have been allowed to vary. The light grey hatched area is where only $\Omega_{m,0}$ varies and $w_0$ and $w_a$ are fixed at the $\Lambda$CDM values of $w_0=-1$ and $w_a=0$ respectively. Panel (b) of Fig.~\ref{fig:combinedplot} illustrates that most of the magnitude discrepancy with the best fitting $\Lambda$CDM comes from the uncertainty in $\Omega_{m,0}$, rather than the choice of dark energy model and uncertainty in $w_0$ and $w_a$. Figures analogous to panel (b) of Fig.~\ref{fig:combinedplot}, but for the JBP, $n3$CPL, and $n7$CPL cosmologies can be found in Appendix \ref{ap:additionalplots}. They likewise show that the majority of the discrepancy with the best fitting $\Lambda$CDM cosmology comes from the uncertainty in $\Omega_{m,0}$, rather than the choice of $w_z$CDM model and uncertainty in $w_0$ and $w_a$.\\

\subsection{Thawing Models}
\label{sec:thawing}
In this subsection the results for the models better suited for fitting an underlying thawing model are presented, namely the results of the CPL, JBP, and $n3$CPL models. In panels (c), (d), and (e) of Fig.~\ref{fig:combinedplot}, $\deltamu$ values are plotted for a representative sample of the parameter values $P=(\Omega_{m,0}, w_0, w_a)$ on the 68\% likelihood contours of the CPL, JBP, and $n3$CPL cosmologies. The overall agreement between the models is remarkable. The extrema of $\deltamu$ for all three models are found at redshift $z \sim 2$, which aligns with the prediction of $\Lambda$CDM that dark energy has negligible influence at larger redshifts. Our results are consistent with the findings of \cite{king2013} who use Fisher matrix analysis to show that a long redshift baseline is important to achieve tight constraints on $w_0$ and $w_a$ for the CPL cosmology, but any measurements beyond a redshift of $z \sim 2$ provide negligible additional constraints compared to that of a lower redshift equivalent.\\

The maximum absolute value of $\deltamu$ when placing no restrictions on the evolution of the equation of state of dark energy is approximately 0.05 mag for all models discussed in this section.  When discarding results where the equation of state enters the phantom regime the lower limits are not strongly affected, but the upper extrema reduces to approximately 0.03 mag.\\\\

\begin{figure}
  \centering\includegraphics[width=\linewidth]{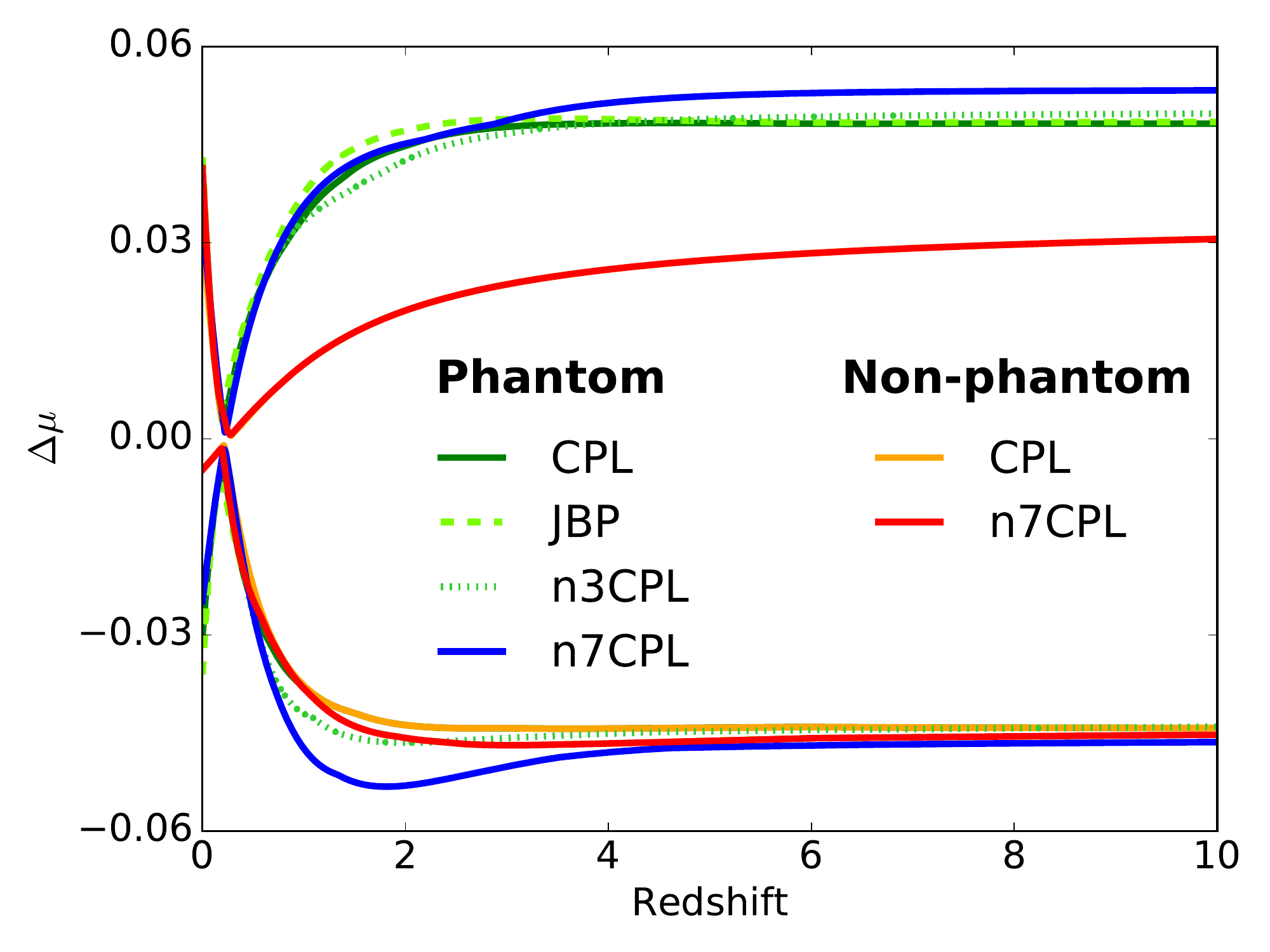}
  \caption{The extrema of $\deltamu$ for the CPL, JBP, $n3$CPL, and $n7$CPL models. The figure illustrates that the models better suited for fitting an underlying thawing dark energy, CPL, JBP and $n3$CPL, all give very similar results. The one model better suited for fitting an underlying freezing dark energy, $n7$CPL, has only slightly larger extrema for $\deltamu$ at high redshift. The red and orange lines show the extrema of the CPL and $n7$CPL excluding phantom models. The non-phantom extrema of the JBP and $n3$CPL models are indistinguishable from those of the CPL model and are therefore not shown.}
  \label{fig:deltamusextrema}
\end{figure}

These results imply that the additional power of high redshift standard candles to discern between $\Lambda$CDM and CPL, JBP, and $n3$CPL cosmology is limited when compared to low redshift standard candles. It also indicates that CPL, JBP, and $n3$CPL cosmology can only deviate slightly from $\Lambda$CDM at large redshifts.\\

\subsection{Freezing Model}
\label{sec:freezing}

In this subsection the results for the model better suited for fitting an underlying freezing cosmology, namely the $n7$CPL model, are presented. In panel (f) of Fig.~\ref{fig:combinedplot} a representative sample of $\deltamu$ values as well as the extrema as dashed lines are plotted for the $n7$CPL cosmology. Overall the similarity to the corresponding plots for the CPL, JBP and $n3$CPL models is strong. In Fig.~\ref{fig:deltamusextrema} the $\deltamu$ extrema for all tested cosmologies are overplotted for comparison. It is apparent that the CPL, JBP, and $n3$CPL models give similar results at both low and high redshift. As expected the $n7$CPL model has the largest extrema at high redshift, but only slightly larger values. This analysis suggests that the conclusion for the $n7$CPL cosmology is similar to that of the CPL, JBP and $n3$CPL cosmologies. Given current constraints they are all unable to deviate significantly from fiducial $\Lambda$CDM at high redshifts. For the $n7$CPL model most of the disagreement with the fiducial $\Lambda$CDM comes from the uncertainty in $\Omega_{m,0}$, rather than the choice of dark energy or uncertainty in $w_0$ and $w_a$, just as was the case with the CPL, JBP, and $n3$CPL models.

\section{Discussion}
\label{sec:discussion}
Our analysis shows that none of the tested cosmologies can deviate significantly from fiducial $\Lambda$CDM cosmology at high redshift, given current cosmological constraints. As a consequence, high redshift ($z\gtrsim 2$) standard candles will not add significant additional constraints over that of low redshift equivalent standard candles, when discerning between $\Lambda$CDM cosmology and CPL, JBP, $n3$CPL, or $n7$CPL cosmology. Since our analysis further shows that the bulk of the disagreement with the fiducial $\Lambda$CDM cosmology for all tested models primarily comes from the uncertainty in $\Omega_{m,0}$, rather than the choice of dark energy model and uncertainty in $w_0$ and $w_a$, we conjecture that if the analysis was to be carried out for any other $w_z$CDM cosmology, it would arrive at results very similar to those of our analysis. The choice of dark energy model seems to have negligible impact on high redshift behaviour, regardless of whether we consider the models better suited for fitting an underlying freezing or thawing cosmology.\\

From our analysis alone it is not possible to conclude that in general no other model better suited for fitting an underlying freezing dark energy could significantly deviate from fiducial $\Lambda$CDM at high redshifts. \cite{Linder2017} investigated a number of freezing dark energy models. Specifically, \cite{Linder2017} produced two parameter functions to calculate observables, such as the dark energy equation of state, for a number of models including the inverse power law (IPL) and supergravity (SUGRA) models. Combined these models span a wide range of possible behaviours for the dark energy equation of state as a function of redshift. \cite{Linder2017} shows that the observables of these complicated models can be approximated with a phenomenological function of two parameters. Varying just these two parameters is sufficient to reproduce the behaviour of these more complicated models with many parameters to within 2.4\% accuracy in the equation of state. The findings of \cite{Linder2017} are not directly transferable to our analysis, but indicate that the complicated behaviour of these dark energy models can be reproduced reliably with just two free parameters. This suggests that even more complicated dark energy models may have very similar behaviour in the equation of state of dark energy, and that our test of the $n7$CPL model may therefore be representative for the much larger family of possible $w_z$CDM models better suited for fitting an underlying freezing dark energy. This indicates that no $w_z$CDM cosmology with only two free parameters for the dark energy redshift behaviour, either thawing or freezing, can deviate significantly from the best fitting $\Lambda$CDM at high redshift.\\

At high redshift additional observational effects beside precision of measurements can impact the usefulness of standard candles. \cite{Holz2005} discuss the effects of gravitational lensing, noting that since the amplification probability distribution is skewed towards deamplification, gravitational lensing will introduce a bias in the measured magnitudes. \cite{Holz2005} find that for a standard candle with intrinsic dispersion of 0.1 mag at redshift 2, gravitational lensing introduces an offset with a mode of 0.03 mag. This effect is of similar size to the maximum offset of 0.05 mag we find for the models tested in this work. This further strengthens the argument that if high redshift standard candles are found to be in disagreement with $\Lambda$CDM cosmology, no $w_z$CDM cosmology will be able to resolve that disagreement.\\

Current or future surveys investing time into measuring high redshift ($z \gtrsim 2$) standard candles could do so for a number of reasons. If the goal is to discern any $w_z$CDM model of the types tested here from a standard $\Lambda$CDM cosmology then the time is best spent observing at redshifts of $z \sim 2$ or lower. However, if the goal is to determine whether the predictions of $\Lambda$CDM hold true at high redshift then going to $z \gtrsim 2$ or higher is recommended, since in this regime there is the added bonus that if $\Lambda$CDM is found to give incorrect predictions, then so will any $w_z$CDM cosmology of the kinds tested in this work.
\section*{Acknowledgements}
The authors would like to thank Eric Linder for useful comments and Tamara Davis for many discussions on the content as well as useful comments. The Dark Cosmology Centre was funded by the Danish National Research Foundation.

\bibliographystyle{mnras}
\bibliography{High_z_standard_candle}
\appendix
\section{Dark Energy Equation of State Parametrisations}
\label{ap:eosgymnastics}
\subsection{$n$CPL}
When working with a $n$CPL cosmology ($w(a) = w_0 + w_a (1-a)^n$), the solution to the continuity equation
\begin{align}
\begin{split}
&\Omega_\Lambda(z) = \Omega_{\Lambda,0}\exp{\left[ \int_{a_0}^a \frac{-3(1+w_{\Lambda}(a))}{a} \mathrm{d} a\right]}\\
&\Omega_\Lambda(z) = \Omega_{\Lambda,0} f(a)
\label{eq:continuity}
\end{split}
\end{align}
for the chosen value of $n$, determines how dark energy evolves with time in the chosen cosmology. Here $f(a)$ is shorthand notation for the function describing the evolution of dark energy with scale factor or redshift. $\Omega_{\Lambda}$ is the energy density of dark energy normalised with the critical energy density, the subscript $0$ indicating the value at the present day. The integral on the right hand side of Eq.~\ref{eq:continuity} is of particular interest, as deriving a general solution would enable directly observing the behaviour of dark energy for the chosen value of $n$ for the $n$CPL cosmology. Using that in the $n$CPL cosmology $w_{n\mathrm{CPL}}(a) = w_0 + w_a (1-a)^n$ and substituting gives
\begin{equation}
f(a)=\int_{a_0}^a \frac{-3(1+w_0 + w_a (1-a)^n)}{a} \mathrm{d} a
\end{equation}
which has the solution
\begin{align}
f(a)=-3\left[ \frac{w_a(1-a)^n \left(\frac{a-1}{a}\right)^{-n}F(n,a)}{n} + (w_0+1)\log{(a)}\right]
\end{align}
with $F(n,a)$ being a power series from the hypergeometric function ${}_2F_1(-n,-n;1-n;\frac{1}{a})$ and using that $a_0=1$.\\

Although the general solution is rather unpleasant to work with, specific cases where $n={1,2,3}$ can be easily solved. The solution for $n=1$ is
\begin{equation}
\Omega_\Lambda(a)_{n=1} = \Omega_{\Lambda,0} a^{-3(1+w_0+w_a)}\exp{(-3w_a(1-a))}.
\end{equation}
For $n=2$
\begin{equation}
\Omega_\Lambda(a)_{n=2} = \Omega_{\Lambda,0} a^{-3(1+w_0+w_a)}\exp{\left(-\frac{3}{2}w_a(1-a)(3-a)\right)}.
\end{equation}
For $n=3$
\begin{align}
\begin{split}
\Omega_\Lambda(a)_{n=3} = &\Omega_{\Lambda,0} a^{-3(1+w_0+w_a)} \times \\
& \exp{\left(-\frac{1}{2}w_a(1-a)(2a^2-7a+11)\right)}.
\end{split}
\end{align}
Inspired by \cite{pantazis2016} the solution for the $n=7$ case is also derived.
\begin{align*}
&\Omega_\Lambda(a)_{n=7} = \Omega_{\Lambda,0} a^{-3(1+w_0+w_a)}\exp{[-\frac{1}{140}w_a(1-a) \times} \\ &{(60a^6 - 430a^5 + 1334a^4 - 2341a^3 + 2559a^2 -1851a + 1089)]}.
\end{align*}

\subsection{JBP}
Similar to the approach in the previous section the continuity equation for the JBP parametrisation, $w(a)_{\mathrm{JBP}} = w_0 + w_a(1-a)a$, is solved yielding
\begin{equation}
\Omega_\Lambda(a)_{\mathrm{JBP}} = \Omega_{\Lambda,0} a^{-3(1+w_0)}\exp{\left(\frac{3}{2}w_a(1-a)^2\right)}.
\end{equation}
\label{ap:parametrisations}
\section{Application of Extreme Value Theorem}
\label{ap:extremevaluetheorem}
When determining the extrema of $\deltamu $ in the space of 68\% likelihood $(\Omega_{m,0}, w_0, w_a)$ values it is helpful to apply the extreme value theorem to state that extrema do exist, and they exist only on the boundary. First, in order for this argument to hold true, we need to prove that $\frac{\partial \deltamu}{\partial \lambda_i} \ne 0$ for all allowed values of one or more of the parameters $\lambda_i \in (\Omega_{m,0}, w_0, w_a)$. {For the following math to be as simple as possible, we will focus on the case where $\lambda_i = w_0$}. First, we define the needed equations. We simplify $\frac{\partial \deltamu}{\partial w_0}$ by noting that
\begin{align}
\begin{split}
\frac{\partial \deltamu}{\partial w_0} &= \frac{\partial (\mu_{w_zCDM}-\mu_{\Lambda CDM})}{\partial w_0} \\
&= \frac{\partial \mu_{w_zCDM}}{\partial w_0} - \frac{\partial \mu_{\Lambda CDM}}{\partial w_0} \\
&= \frac{\partial \mu_{w_zCDM}}{\partial w_0} - 0 \\
&= \frac{\partial \mu_{w_zCDM}}{\partial w_0} \\
&= \frac{\partial \mu}{\partial w_0} 
\end{split}
\end{align}
where for simplicity we defined $\mu = \mu_{w_zCDM}$ in the final line. Following \cite{king2013} the equations needed now are
\begin{equation}
\frac{\partial \mu}{\partial w_0} = \frac{5}{D^{'}_M \ln{(10)}}\frac{\partial D^{'}_M}{\partial w_0},
\end{equation}
where $D^{'}_M$ is the dimensionless tangential comoving distance $D^{'}_M=(H_0/c)D_M$. In the limit where the curvature term $\Omega_k$ approaches zero, i.e. a cosmology that is close to flat, this gives
\begin{align}
&\lim_{\Omega_k \rightarrow 0} \frac{\partial D^{'}_M}{\partial w_0} = \frac{\partial \chi^{'}}{\partial w_0}, \\
& \frac{\partial \chi^{'}}{\partial w_0} = -\frac{1}{2}\int_0^z \frac{1}{E(z^{'})^3} \frac{\partial \mathcal{E}(z^{'})}{\partial w_0} \mathrm{d}z^{'}, \\
&\frac{\partial \mathcal{E}(z)}{\partial w_0} = \Omega_x \frac{\partial f(z)}{\partial w_0}, \\
&\frac{\partial \mathcal{E}(z)}{\partial w_0} = 3\Omega_x f(z) \ln{(1+z)},
\end{align}
and finally
\begin{align}
& \mathcal{E}(z) = H(z)^2/H_0^2 = E(z)^2, \\
\begin{split}
 E(z)^2 &= \Omega_{m,0}(1+z)^3 + \Omega_k(1+z)^2 + \\
& \Omega_{\Lambda,0} \exp{\left(3\int_0^{z} [1+w(z^{'})]\frac{\mathrm{d}z^{'}}{1+z^{'}}\right)}.
\end{split}
\end{align}
If we show that all terms added in the above equations have the same sign, that is they are either positive or negative for all values of $z$, it follows that they never cancel out and will therefore never sum to zero. Noting that
\begin{enumerate}
\item $E(z)$ is for flat cosmologies only ever positive,
\item$f(z) > 0$ for $n$CPL and JBP parametrisations,
\item and $z > 0$
\end{enumerate}
it follows that $\frac{\partial \deltamu}{\partial w_0}$ is only ever negative, if we restrict ourselves to work with flat cosmologies. Applying the extreme value theorem, it is apparent that there are no critical points, so the extreme values of $\deltamu$ exist on the boundary. This argument reduces the dimensionality of the problem by one from three to two, saving a large amount of computation time in searching for the extreme values of $\deltamu$.
\section{Data Analysis}
\label{ap:datanalysis}
In this work CosmoMC \citep{Lewis2002, Lewis2013}, utilising CAMB \citep{Lewis2000,Howlett2012}, PICO \citep{Fendt2006}, and CMBFAST \citep{Seljak1996, Zaldarriaga1997} was used extensively. The datasets used include observations of the cosmic microwave background from the Planck Collaboration \citep{PlanckCollaboration2015, PlanckCollaboration2015a, PlanckCollaboration2015b} and BICEP-Planck \citep{BICEP2/Keck2015}, observations of baryon acoustic oscillations from SDSS-III \citep{Anderson2013}, 6dF \citep{Beutler2012}, MGS \citep{Blake2011, Padmanabhan2012, Anderson2013a, Ross2015} as well as supernova data from the Joint Light-Curve Analysis \citep{Betoule2014}. For analysis of the output chains of CosmoMC this work made use of Astropy, a community-developed core Python package for Astronomy \citep{TheAstropyCollaboration2013}. For details on how the output chains of CosmoMC were analysed and plots of the results were produced see the GitHub repository at \url{https://github.com/per-andersen/Deltamu}.
\section{Additional Plots}
\label{ap:additionalplots}
\begin{figure}
  \centering\includegraphics[width=\linewidth]{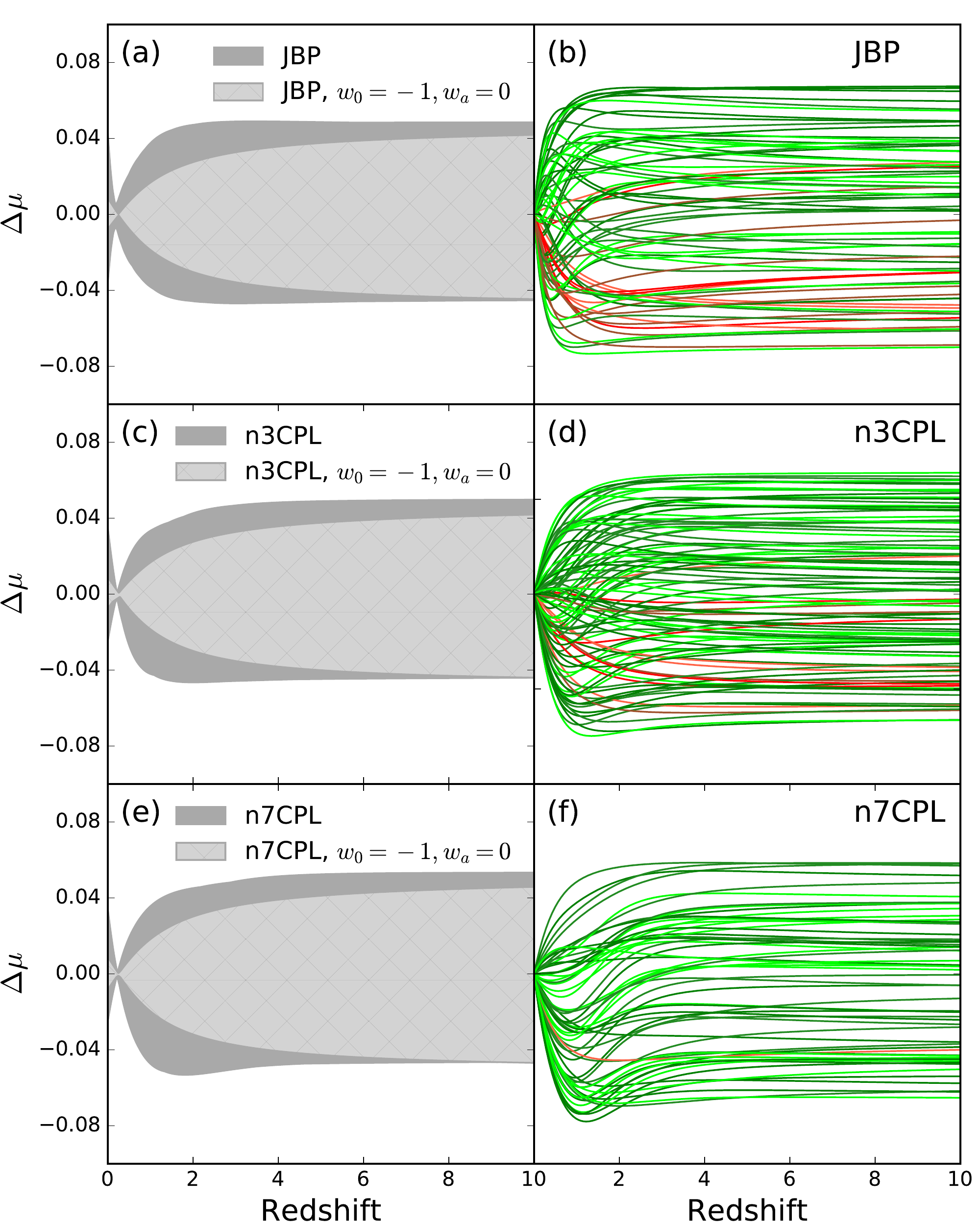}
  \caption{Plots analogous to panels (a) and (b) of Fig.~\ref{fig:combinedplot}, but instead of the CPL model for the JBP, $n3$CPL, and $n7$CPL models.}
  \label{fig:additionalplots}
\end{figure}
In Fig.~\ref{fig:additionalplots} plots analogous to panels (a) and (b) of Fig.~\ref{fig:combinedplot} are shown for the JBP, $n3$CPL, and $n7$CPL models. First, in panel (a), (c), and (e) it is shown that the largest contribution to the extrema of $\deltamu$ comes from the uncertainty in $\Omega_{m,0}$, rather than the choice of dark energy model and uncertainty in $w_0$ and $w_a$, just like for the CPL model. In panels (b), (d), and (f) the effect of $K$ on the JBP, $n3$CPL, and $n7$CPL models is shown to be similar to that on the $\deltamu$ of the CPL model; not including $K$ removes the scatter around $z=0$ and widens the distribution of $\deltamu$ somewhat, increasing the extrema of $\deltamu$ at high redshifts.
\end{document}